\documentclass[pra,showpacs,twocolumn,superscriptaddress,floatfix]{revtex4}
\usepackage{times,graphicx,amsmath,bbm,amssymb}
 
 \newcommand{\sfy}{{\sf y}}

\newcommand{\bmsigma}{\boldsymbol \sigma}

\newcommand{\bmX}{\boldsymbol X} \newcommand{\bmA}{\boldsymbol A}
\newcommand{\bmM}{\boldsymbol M}\newcommand{\bmR}{\boldsymbol R}
 
 \newcommand{\bmLambda}{{\boldsymbol \Lambda}}
 \newcommand{\sfx}{{\sf x}}

\begin{document}
\title{Entanglement oscillations in non-Markovian quantum channels}
\author{Sabrina Maniscalco}
\email{sabrina.maniscalco@utu.fi} \affiliation{Department of
Physics, University of Turku, FI-20014 Turku, Finland}
\author{Stefano Olivares}
\email{stefano.olivares@mi.infn.it}
\affiliation{Dipartimento di Fisica dell'Universit\`a di Milano, I-20133, Italia}
\author{Matteo G. A. Paris}
\email{matteo.paris@fisica.unimi.it}
\affiliation{Dipartimento di Fisica dell'Universit\`a di Milano, I-20133, Italia}
\affiliation{Institute for Scientific Interchange Foundation, I-10133, Torino,
Italia}
\date{\today}
\begin{abstract}
We study the non-Markovian dynamics of a two-mode bosonic system
interacting with two uncorrelated thermal bosonic reservoirs.  We
present the solution to the exact microscopic Master equation in terms
of the quantum characteristic function and study in details the dynamics
of entanglement for bipartite Gaussian states. In particular, we analyze
the effects of short-time system-reservoir correlations on the
separability thresholds and show that the relevant parameter is the
reservoir spectral density.  If the frequencies of the involved modes
are within the reservoir spectral density entanglement persists for a
longer time than in a Markovian channel.  On the other hand, when the
reservoir spectrum is out of resonance short-time correlations lead to a
faster decoherence and to the appearance of entanglement oscillations.
\end{abstract}
\maketitle
The loss of coherence, or decoherence, within a quantum system is
due to the interactions of the system with the surrounding
environment.  Decoherence processes cannot be ignored, for they
are the main obstacle to the full realization of quantum
information processing. The dynamics of open quantum systems,
however, may be rather involved, mostly due to the complex
structure of the environment interacting with the quantum system.
Therefore, in order to describe the dynamics of the system of
interest, some approximations are often made, leading to the
derivation of a Master equation for the reduced density matrix.
The most relevant approximations are the weak coupling or {\em
Born approximation}~\cite{petruccionebook,Weissbook}, assuming
that the coupling between the system and the reservoir is small
enough to justify a perturbative approach, and the {\em Markov
approximation}~\cite{petruccionebook,Weissbook}, which amounts to
neglect short-time correlations between the system and the
reservoir. There are few cases, however, where an exact analytic
description of the dynamics is possible. Two relevant examples
are the quantum Brownian motion
(QBM)~\cite{petruccionebook,Weissbook,Zurek,HuPazZhang} and the
case of a two-level atom interacting with a thermal reservoir
with Lorentzian spectral
density~\cite{petruccionebook,Garraway97}.
\par
Entanglement in continuous variable (CV) quantum channels has 
attracted much interest in recent years, due to potential 
improvement in the channel capacity \cite{P2}.
A realistic analysis of CV channels must take into account 
decoherence and dissipation phenomena and, in fact, there has 
been an increasing interest in the description of noisy CV quantum
channels \cite{SeraRev}. Most of the theoretical descriptions,
however, rely on Born and/or Markov approximations and only very
recently some phenomenological models of non-Markovian quantum
channels have been proposed \cite{Ban,KimJMO}. Non-Markovian
effects are crucial, e.g., for high-speed quantum communication
where the characteristic time scales become comparable with the
reservoir correlation time. Moreover, when the system interacts
with a structured reservoir, e.g., for quantum channels embedded in
solid-state devices, memory effects are typically non negligible.
In these cases the dynamics can be substantially different from
the Markovian one.
\par
In this communication we focus attention on the dynamics of a
two-mode bosonic quantum system propagating in a noisy bosonic quantum
channel, i.e., interacting with two uncorrelated bosonic thermal
reservoir.
Such a system may be seen as made by
two quantum Brownian oscillators and therefore it is possible to
exactly describe the system dynamics.  Using this analogy, we
generalize the single mode QBM solution to the bimodal case and
analyze in details the entanglement dynamics of bipartite Gaussian
state.  We present an exact approach based on a microscopic model and,
starting from the exact solution, we discuss in details the genuine
non-Markovian effects on the entanglement dynamics for the relevant
class of bipartite Gaussian states. In particular we analyze the
effects of short-time system-reservoir correlations on the
separability thresholds.  The exact results are compared with those
obtained within the Markovian approximation, and few relevant
situations in which the system-reservoir correlations give rise, for
short times, to entanglement oscillations are addressed.  We stress that our
approach describes the effects of quantum noise for two-mode quantum
systems starting from a microscopic description of the system and the
reservoir.
\par
The exact Master equation describing a quantum harmonic oscillator
interacting with a bosonic reservoir in thermal
equilibrium has been derived for the first time in
Ref.~\cite{HuPazZhang} and it is usually referred to as
Hu-Paz-Zhang Master equation. The original derivation is based on
the influence functional-path integral formalism though a simpler
derivation based on the time-convolutionless projection operator
technique \cite{petruccionebook} may be also used [see, e.g.,
Ref.~\cite{Intravaia1}]. Assuming that the bimodal field interacts
bilinearly with two identical uncorrelated
bosonic thermal reservoirs the Master equation for the reduced density
matrix $\varrho(t)$ of the field is given by
\begin{align}
 \dot\varrho(t) =&  \sum_k \Big\{ \frac{1}{i \hbar} [H_k^0,\varrho (t)]
 -   \Delta(t) [X_k,[X_k,\varrho(t)]]  \nonumber \\
 &+ \Pi(t) [X_k,[P_k,\varrho(t)]]+ \frac{i}{2} r(t)
 [X_k^2,\varrho (t)] \nonumber \\
 &-  i \gamma(t) [X_k,\{P_k,\varrho\}]\Big\}, \label{QBMme}
\end{align}
where  
$$X_k = \frac{1}{\sqrt{2}} \left( a_k + a_k^{\dag}\right) \quad 
P_k=\frac{i}{\sqrt{2}} \left( a_k^{\dag}- a_k\right)$$ 
are the dimensionless quadrature operators, $[a_k, a^{\dag}_k]=1$
($k=1,2$) being the mode operators of the two oscillators, and
$H_k^0= \hbar \omega_0 (a^{\dag}_k a_k+1/2)$. This Master
equation, being exact, describes also the non-Markovian
system-reservoir correlations due to the finite correlation time
of the reservoir. In contrast to other non-Markovian dynamical
models~\cite{Barnett01}, Eq.~(\ref{QBMme}) is local in time,
i.e.~it does not contain memory integrals. All the non-Markovian
character of the system is contained in the time dependent
coefficients, $\Delta (t)$, $\Pi(t)$, $r(t)$ and $\gamma(t)$,
appearing in the Master equation. These coefficients depend
only on the reservoir spectral density, i.e.~on
the microscopic effective coupling strength between the system
oscillator and the oscillators of the reservoir. The coefficient
$r(t)$ describes a time dependent frequency shift, $\gamma(t)$ is
the damping coefficient, $\Delta(t)$ and $\Pi(t)$ are the normal
and the anomalous diffusion coefficients, respectively
\cite{petruccionebook,Zurek}.
It is worth
underlining that the Master equation given by Eq.~(\ref{QBMme}),
is valid for general forms of the reservoir spectral density
$J(\omega)$ and any temperature $T$.
\par
The solution of Eq.~(\ref{QBMme}) can be obtained generalizing to
two-modes the method of solution of the Hu-Paz-Zhang Master
equation. Several approaches have been proposed in the literature
to solve the Hu-Paz-Zhang Master
equation~\cite{solHPZ,PRAsolanalitica,misbelief}. The one
developed in Refs.~\cite{PRAsolanalitica,misbelief} is based on
the symmetrically ordered quantum characteristic function (QCF) 
$\chi_t(\xi)$ at time $t$. Extending this
approach to the case of a bimodal field, in the interaction
picture with respect to $\sum_{k=1,2}^{}H_k^0$ we obtain the
following solution
\begin{eqnarray}\label{evolution}
\chi_t (\bmLambda)= \exp \left\{-  \bmLambda^T \bmA_t
\:\bmLambda\right\} \chi_0 (e^{-\Gamma (t) }\bmLambda),
\label{chit2mode}
\end{eqnarray}
where we indicate with $\chi_0 (\bmLambda)$ the characteristic
function at $t=0$, and where $\bmLambda =
(\sfx_1,\sfy_1,\sfx_2,\sfy_2)^{T}$, $(\cdots)^{T}$ being the
transposition operation.  The bipartite QCF is defined as
\begin{eqnarray}
\chi(\bmLambda)={\rm Tr} \left[e^{\xi_1 a_1^{\dag}-\xi_1^*
a_1} e^{\xi_2 a_2^{\dag}-\xi_2^* a_2} \:\varrho \right],
\end{eqnarray}
with $\xi_k = \frac{1}{\sqrt{2}} (\sfx_k + i \sfy_k)$, and
$k=1,2$. From the QCF one evaluates the moments of the field
\begin{multline}
\langle a_1^{\dag k}a_2^{\dag l} a_1^m a_2^n \rangle =
(-)^{m+n} \partial_{\xi_1}^k
\partial_{\xi_2}^l
\partial_{\xi_1^*}^m
\partial_{\xi_2^*}^n
e^{\frac12(|\xi_1|^2+
|\xi_2|^2)} \\
\times \chi(\xi_1,\xi_2)|_{\xi_1=0,\xi_2=0}
\nonumber \:,
\end{multline}
and, in turn, the time evolution of the covariance matrix, thus
quantifying the entanglement between the two modes.
In Eq.~(\ref{chit2mode}) the matrix $\bmA_t$ is given by~
\cite{misbelief}
\begin{eqnarray}
\bmA_t = e^{-\Gamma (t)} \int_0^t \!\!ds\: e^{\Gamma(s)}
\bmR^{T}(t,s)\:\bmM (s)\: \bmR(t,s),
\end{eqnarray}
where the matrix $\bmR(t,s)$ contains rapidly oscillating terms
and $\bmM (t)$ is given by
\begin{equation}
\bmM (t) = \frac12 \left(
\begin{array}{cc}
2 \Delta(t) & -\Pi(t)  \\
-\Pi(t) & 0
\end{array}
\right).
\end{equation}
It is worth recalling that the time dependent coefficients
$\Delta(t)$, $\Pi(t)$ and $\gamma(t)$ depend only on the
spectral density of the reservoir and can be expressed as power
series in the system-reservoir coupling constant $\alpha$. 
Finally, $\Gamma(t)$ takes the form
\begin{eqnarray}
\Gamma(t)=2\int_0^t\!\! ds\: \gamma(s)\:. \label{Gamma}
\end{eqnarray}
Once the spectral characteristics of the environment and the
initial state of the field have been specified, then
Eq.~(\ref{chit2mode}) fully characterizes the
dynamics of the system without approximations.
\par
We now consider a specific example of interest for CV
quantum information and communication, namely the case of
a bipartite Gaussian state of the form
\begin{equation}
\chi_0(\bmLambda) = \exp\left\{ -\frac12 \bmLambda^{T} \bmsigma_0
\bmLambda -i \bmLambda^{T}\, \overline{\bmX}_{\rm in} \right\},
\end{equation}
where
\begin{equation}
\bmsigma_0 = \left(
\begin{array}{c|c}
{\boldsymbol A} &{\boldsymbol C}\\
\hline
{\boldsymbol C} &{\boldsymbol B}
\end{array}
\right)
\end{equation}
with ${\boldsymbol A}={\rm Diag}(a,a)$, ${\boldsymbol B}={\rm
Diag}(b,b)$, $a,b>0$, and ${\boldsymbol C}={\rm Diag}(c_1,c_2)$
is the covariance matrix in the canonical form and
$\overline{\bmX}_{\rm in} = {\rm
Tr}[\varrho (t)\: (X_1, P_1, X_2, P_2)^{T}]$ the vector of mean values.
For optical bimodal fields the solution of the Master equation can
be simplified invoking the secular approximation. It has been
shown that, in this case, the time evolution depends only on the
diffusion coefficient $\Delta (t)$ and on the dissipation
coefficient
$\gamma(t)$~\cite{Intravaia1,PRAsolanalitica,misbelief}. The
expression for the QCF at time $t$ becomes
\begin{equation}\label{chit}
\chi_t(\bmLambda) = \exp\left\{ -\frac12 \bmLambda^{T} \bmsigma_t
\bmLambda -i \bmLambda^{T}\, \overline{\bmX}_{\rm t} \right\}
\end{equation}
where the evolved
mean value and covariance matrix are given by
\begin{align}
\overline{\bmX}_t &=e^{-\Gamma(t)/2}(R \oplus R)^{T}\,
\overline{\bmX}_{\rm in} \:, \label{Xevolved} \\
\bmsigma_t &= e^{-\Gamma(t)} (R \oplus R)^{T} \bmsigma_0 (R
\oplus R) + \mbox{$\frac12$}\Delta_{\Gamma}(t) \mathbbm{1}\:,
\label{sigmaevolved}
\end{align}
$\mathbbm{1}$ being the $4\times 4$ identity matrix and
$R$ the rotation matrix
\begin{equation}
R = \left(
\begin{array}{cc}
\cos \omega_0 t & \sin \omega_0 t \\
-\sin \omega_0 t & \cos \omega_0 t
\end{array}
\right)\,.
\end{equation}
Finally, the time dependent coefficient $\Delta_{\Gamma} (t)$
appearing in Eq.~(\ref{sigmaevolved}) is given by
\begin{eqnarray}
\Delta_{\Gamma}(t) &=& e^{-\Gamma(t)}\int_0^t \!\! ds\:
e^{\Gamma(s)}\Delta(s) \label{DeltaGamma}.
\end{eqnarray}
The map expressed by Eqs. (\ref{Xevolved}) and (\ref{sigmaevolved})
is Gaussian, i.e.  it maintains the Gaussian character of the input
state. As a consequence, separability after interactions may be checked
by positivity of the partially transposed density matrix (PPT condition)
\cite{PPT:peres,PPT:simon}, which, in terms of the covariance matrix,
reads as follows
\begin{equation}\label{sep:cond}
\bmsigma_t + \frac{i}{2} {\boldsymbol \Omega} \geq 0\,,
\end{equation}
where ${\boldsymbol \Omega} = {\boldsymbol \omega} \oplus
{\boldsymbol \omega}^{T}$, with
\begin{equation}
{\boldsymbol \omega} = \left(
\begin{array}{cc}
0 & 1 \\
-1 & 0
\end{array}
\right)\,.
\end{equation}
Explicitly, for the evolved state of Eq. (\ref{chit}), the
PPT condition may be written as
\begin{multline}
\left[ (a+b) e^{-\Gamma(t)} + \Delta_{\Gamma}(t) \right]^2 \\
- e^{-\Gamma(t)} \left\{ \left[ (a-b)^2 + 2(c_1^2+c_2^2) \right]
e^{-\Gamma(t)} \right. \\ \left. - 2 \sqrt{(a-b)^2 + (c_1 - c_2)^2
+ (c_1^2 - c_2^2)^2 e^{-2\Gamma(t)}} \right\} \ge 1\,.
\label{PPTgen}
\end{multline}
In the case of a twin-beam (TWB) state of radiation, i.e.,
$$a=b=\frac12 \cosh (2r) \quad c_1 = -c_2 = 
\frac12 \sinh (2r)\:,$$ 
$r>0$, the last inequality reduces to \cite{bech:JPA:04}
\begin{equation}
S(t)=e^{-2r} e^{-\Gamma(t)} + \Delta_{\Gamma}(t) -1 \ge 0\,.
\label{ft}
\end{equation}
Notice that, since in the case of a Gaussian state there is no
{\em bound entanglement},  the sign of $S(t)$ is strictly connected
to the separability of the state: when $S(t)\ge 0$ the state is
separable and, for this reason, $S(t)$ is called  {\em separability
function}.
At $t=0$, $\Gamma(0)=\Delta_{\Gamma}(0)=0$, hence
$S(0)=e^{-2r}-1$, i.e. the initial state is
an entangled state. As time passes, due to the
interaction with
the external environment, the entanglement between the modes
deteriorates and the separability function $S(t)$ vanishes.  The
time $t_{\rm s}$ at which $S(t_{\rm s})=0$, i.e. when the
entanglement between the two modes is lost, is referred to as the
separability threshold or separability time.  Of course, $t_{\rm s}$
is a function of both the system and the
reservoir parameters. In the Markovian approximation, for
example, $t_{\rm s}$ has been shown to become infinite
for a zero $T$ reservoir \cite{SepM}.
\par
In order to study in more detail the entanglement dynamics
in a non-Markovian channel, we need
to specify the spectral density of the reservoir. We focus on the
Ohmic reservoir with Lorentz-Drude regularization, whose spectral
density is \cite{Weissbook}
$$J(\omega)= \frac{\omega}{\pi}
\frac{\omega_c^2}{\omega_c^2+\omega^2}\:,$$ where $\omega_c$ is
the cutoff frequency. This is one of the most  studied models
of bosonic environments since it leads to a friction force
proportional to velocity, which is typical of dissipative systems
in several physical contexts. In this model, the reservoir
correlation time is given by the inverse of the cutoff frequency,
i.e. $\tau_R=1/\omega_c$. For times $t \le \tau_R$ the system
dynamics cannot be approximated by the Markovian dynamics, as we
will show in the following. The expressions for the relevant time
dependent coefficients, up to second order in the system-reservoir
coupling constant, are given by~\cite{HuPazZhang,Maniscalco04b}
\begin{align}
\Delta(t) &= \int_0^t \!\! d\tau\,\kappa(\tau) \cos (\omega_0 \tau)
\label{delta} \\
\gamma(t) &= \int_0^t \!\! d\tau\,\mu(\tau) \sin (\omega_0 \tau)
\label{gamma}
\end{align}
with
\begin{align}
\kappa(\tau)&= \alpha^2 \int_0^{\infty}\!\! d\omega\, J(\omega) \coth [\hbar
\omega /2 k_B T] \cos(\omega \tau ) \label{kappa} \\
\mu(\tau) &=  \alpha^2 \int_0^{\infty}\!\! d\omega J(\omega)  \sin(\omega
\tau )\label{mu}
\end{align}
being the noise and the dissipation kernels,
respectively~\cite{petruccionebook}.
We note that, for high reservoir temperatures $T$, $\Delta(t)\gg
\gamma(t)$. Inserting the expression of $J(\omega)$ into
Eqs.~(\ref{delta}) and (\ref{gamma}) one gets analytic expressions
for both of the coefficients for the high $T$ case. For generic
temperatures only $\gamma(t)$, which is independent of
temperature, has a simple analytic form, while the time integral
in $\Delta(t)$ has to be calculated numerically [see, e.g.,
Ref.~\cite{Maniscalco04b} for the analytic expression of the
diffusion and dissipation coefficients].
We begin our analysis by noting that, for times $t \gg \tau_R$, the
coefficients $\Delta(t)$ and $\gamma(t)$ can be approximated by
their Markovian stationary values $\Delta_M=\Delta(t \rightarrow
\infty )$ and $\gamma_M = \gamma(t\rightarrow\infty)$ [see
Appendix B of Ref.~\cite{Maniscalco04b}]. Eqs.~(\ref{Gamma})
and (\ref{DeltaGamma}), hence, become
\begin{align}
\Gamma(t) &= \gamma_M t \label{gammaM}\\
\Delta_{\Gamma}(t) &= [2 N(\omega_0) + 1](1-e^{-\gamma_M t})
\label{deltaM}
\end{align}
where $N(\omega_0) = (e^{\hbar \omega_0/k_B T}-1)^{-1}$ is the
average number of thermal photons. Inserting
Eqs.~(\ref{gammaM}) and (\ref{deltaM}) into Eq.~(\ref{ft}), one can
easily see that the separability time coincide with that predicted
by the Markovian theory \cite{SepM,bech:JPA:04}, i.e
$$ t_s = \frac1{\gamma_M}
\log \left(1 + \frac{1-e^{-2r}}{2 N(\omega_0)}\right)\:.$$
Notice that, as mentioned above, if $N(\omega_0) \to 0$
then $t_{\rm s} \to \infty$.
\par
\begin{figure}[h]
\begin{tabular}{c}
\includegraphics[width=0.45\textwidth]{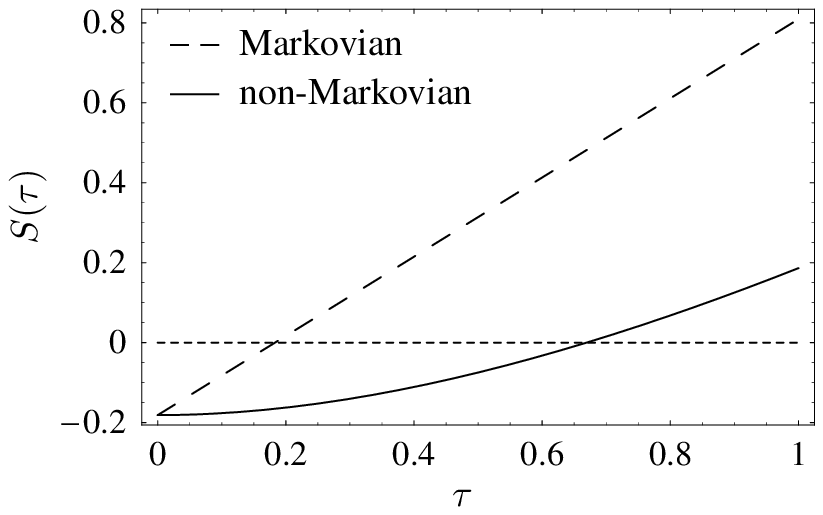} \\
\includegraphics[width=0.45\textwidth]{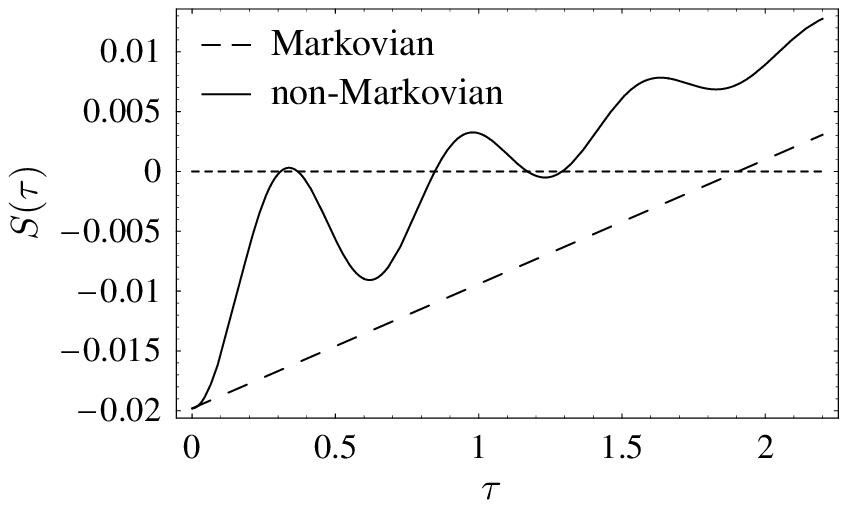}
\end{tabular}
\caption{\label{fig:1}
Separability function $S(\tau)$ for twin-beam state in
Markovian and non-Markovian channels for a high $T$ reservoir.
Plots refer to parameters $k_B T/ \hbar \omega_c = 100 $,
$\alpha^2=0.01$, $r=0.1$, and (Top): $x=10$; (Bottom):
$x=0.01$.
The top plot refers to a situation in which the frequencies of 
the involved modes overlap with the reservoir spectral density 
($x \gg 1$). Here , the more accurate non-Markovian estimation
predicts that the entanglement persists for a longer time. This 
conclusion holds for any value of the
reservoir temperature, provided the high $T$ condition is
satisfied, and as long as $x \gg 1$.
The bottom plot refers to a situation with the reservoir spectrum
out of resonance with respect to the mode frequencies, i.e. when 
$x \ll 1$. Here, the environment causes a faster loss
of entanglement compared to the Markovian prediction, as well as 
to the appearance of entanglement oscillations.}
\end{figure}

We now look at the deviation from the Markovian value of the
separability time for times $t\le \tau_R$. In this time interval
the expression for the separability condition (\ref{ft}) reduces to
\begin{align}
S(t) \simeq e^{-2r} \left( 1-\int_0^t \!\!ds
\:\gamma(s)\right)+\int_0^t \!\!ds\: \Delta(s) -1 \ge 0.
\end{align}
We consider the high $T$ case. Using the analytic expressions of
$\Delta(t)$ and $\gamma(t)$, we get
\begin{multline}
S(\tau)=\alpha^2 \frac{k_B T}{\hbar \omega_c} \frac{x^2}{1+x^2}
\left\{ \tau - \frac{x^2-1}{x^2+1} \left[ 1-e^{-\tau}\cos (\tau/x)
\right] \right. \\
- \left. 2\frac{x}{x^2+1} e^{-\tau} \sin (\tau/x) \right\} +
e^{-2r}-1\:, \label{fhight}
\end{multline}
with $\tau = \omega_c t$, and $x= \omega_c / \omega_0$. In
Fig.~\ref{fig:1} we report the separability function $S(t)$,
and we compare it with the corresponding Markovian approximation
\begin{align}
S_M(\tau) =  
\frac{\tau}{2}
\frac{\alpha^2 x}{1+x^2}
\left(1+2 x \frac{k_B T}{\hbar \omega_c} \right)
+ e^{-2r}-1
\:.
\end{align}
\par
The case considered in the top plot corresponds to a situation in
which the frequencies of the involved modes overlap with the
reservoir spectral density ($x \gg 1$). The Markovian prediction
for the separability time, in this case, gives a smaller value
than the one predicted by our non-Markovian theory. More precisely
we have $\tau_{\rm s} \simeq 0.18 \le \tau_{\rm s} \simeq
0.65$. Therefore, the more accurate non-Markovian estimation
predicts that the entanglement persists for a longer time. It is
easy to show that this conclusion holds for any value of the
reservoir temperature, provided the high $T$ condition is
satisfied, and as long as $x \gg 1$.
\par
The situation changes for the case in which the reservoir spectrum
is ``out of resonance''  or ``detuned''  with respect to the
frequency of the TWB, i.e. when $x \ll 1$. As shown in Fig.~1
(bottom plot), in this case the environment causes a faster loss
of entanglement compared to the Markovian prediction. A
careful analysis of Eq.~(\ref{ft}) shows that this feature can be
traced back to the short time behavior of the diffusion
coefficient $\Delta(t)$, describing environment induced
decoherence~\cite{HuPazZhang}. For $x \gg 1$, indeed, $\Delta(t)
\ge \Delta_M$, with $\Delta_M = \Delta(t\rightarrow \infty)$ the
Markovian value. Therefore, in this case, the system experiences a
weaker decoherence than the one predicted by the Markovian
approximation and, as a consequence, the entanglement between the
modes persists longer. On the contrary, for $x \ll 1$, $\Delta(t)$
quickly assumes values higher than $\Delta_M$. The system hence is
subjected to a stronger initial decoherence which quickly destroys
the quantum entanglement between the two modes. In Fig.~1 we see
that, for $x \ll 1$, the separability function displays
entanglement oscillations, a typical non-Markovian feature.
Similar oscillations have been extensively studied for the case of
the one-mode harmonic quantum Brownian model~\cite{oscillations}.
It is worth noting, however, that here we look at quantum
correlations between two modes whose dynamics cannot be directly
deduced from the dissipative dynamics of a single mode bosonic
quantum channel.
\par
In conclusion, we have analyzed the dynamics of bipartite Gaussian
states in a non-Markovian noisy channel. We found that when
the mode frequencies are within the reservoir spectral density
entanglement persists for a longer time than in a Markovian channel.
On the other hand, when the reservoir spectrum is out of resonance
than short-time correlations lead to a faster decoherence and to
the appearance of entanglement oscillations. Our analysis indicates
that non Markovian features of system-reservoir interaction may be
relevant in quantum information processing and should be taken into account
in the design of any quantum communication channel.
\par
This work has been supported by MIUR through the project
PRIN-2005024254-002. S.M. acknowledges financial support from the
Academy of Finland (projects 108699,115682) and from the Magnus
Ehrnrooth Foundation.

\end{document}